# Authority Signals in Claude AI Health Citations: A Descriptive Analysis Using the Authority Signals Framework

Erin T. Jacques[1], Erela Datuowei[2], Elizabeth Quaye[3], Corey H. Basch[4], Arijit Chatterjee[1], Juanita Davis[1]

[1]Department of Health and Human Performance, York College, The City University of New York, 94-20 Guy R. Brewer Blvd. Jamaica, NY 11451

[2]Department of Health Studies & Applied Educational Psychology, Teachers College, Columbia University, 525 West 120th Street, New York, NY 10027

[3]Department of Accounting and Finance, York College, The City University of New York, 94-20 Guy R. Brewer Blvd. Jamaica, NY 11451

[4]Department of Public Health, William Paterson University, 300 Pompton Road Wayne, NJ, 07470

**Corresponding Author: ejacques1@york.cuny.edu**

## Abstract

This study seeks to determine the authority signals used by Anthropic's Claude AI in its presentation of sources when answering consumer health questions. While there exists a great deal of discourse around the quality of health citations that LLMs produce, there is limited information on the integrity of the sources the citations originate from, and to what extent the sources are, from what health professionals would consider, credible sources. This descriptive cross-sectional study used data from HealthSearchQA, which contains 3,172 consumer health questions curated by Google Research. After exclusions, a final dataset of 3,075 questions yielding 10,038 citations was analyzed. The Authority Signals Framework (Jacques et al., 2026) was applied to examine 10 authority signals across four domains for a disproportionate stratified sample of 542 sources. Established institutional sources accounted for 97.8% of all citations (n = 9,818). Medical Institutions were the most frequently cited organization type (36.5%), followed by Government Resources (31.6%) and Professional Associations (28.4%). Commercial Health Information comprised 2.2% (n = 220). The top 10 organizations accounted for 57.8% of all citations, with Mayo Clinic alone representing 24.7%. Among commercial sources in the focused sample, 86.4% displayed medical review statements, 82.5% used schema markup, and 71.8% had comprehensive content, while traditional institutional sources appeared in Claude's citations with or without these same markers. As Anthropic positions Claude for HIPAA-ready healthcare applications, these findings establish a baseline for Claude's citation behavior and demonstrate the utility of the Authority Signals Framework as a tool for ongoing, cross-platform evaluation of AI-mediated health information.

## 1. Introduction

Expanding on prior research, this study seeks to determine the authority signals used by Anthropic's Claude AI in its presentation of sources when answering consumer health questions (Jacques et al., 2026). Originally formed in 2021 as a safer and more transparent response to OpenAI's ChatGPT, Claude AI falls under what is termed as Constitutional AI, artificial intelligence trained on an explicit set of principles aligned with human values (Fortune Editors, 2023). Less than five years later, Claude AI now boasts ownership of 32% of the Enterprise Large Language Model (LLM) market and 42% of the LLM code generation market (Fortune Editors, 2023; Szkutak, 2025). This growth has extended into the healthcare sphere as well. Announced in early January 2026, Claude for Healthcare is a new set of tools that will allow healthcare providers, payers, and consumers to use Claude for medical purposes through HIPAA-ready products (Anthropic, 2026). Providers and payers will have access to connectors that can directly access Centers for Medicare & Medicaid Services (CMS) Coverage Database, the tenth revision of the International Classification of Diseases (ICD-10), and National Provider Identifier Registry. HIPAA-compliant organizations will also have access to new packages that allow for the development of healthcare data exchanges via Fast Healthcare Interoperability Resources (FHIR) as well as sample prior authorization review (Anthropic, 2026).

This new tool underscores the continued large-scale adoption of Large Language Models (LLMs) in healthcare in new and novel contexts beyond its current uses. Considering this exponential growth and Anthropic's positioning of Claude AI as a more reliable alternative to ChatGPT, it is important to probe the validity of the information it transmits to its users, both providers and consumers, as this can negatively impact health outcomes (Jin et al., 2024). A recent analysis found that half of all AI-generated search results lacked citations. Of the results that did include citations, only 75% actually support the claim (Linardon et al., 2025). A 2025 study estimated that the percentage of erroneous citations may be even higher, between 50% and 90% (Wu et al., 2025). Linardon et al. (2025) also found that reliability of LLM health citations decreased as health disorders became less visible and more specialized. While these citation hallucinations may occur as a result of vague or structurally misleading prompts by the user, they may also occur due to limitations in parametric knowledge, architectural biases, or inference-time sampling strategies (Anh-Hoang et al., 2025). Yet, while these studies examine whether LLM citations are accurate, less is known about the characteristics of the sources themselves, specifically, whether cited sources reflect what health professionals would consider credible.

While there exists a great deal of discourse around the quality of health citations that LLMs produce, there is limited information on the credibility of the sources that are present, and to what extent the sources are from what health professionals would consider credible sources. Traditional markers of credible health information sources include government agencies, medical institutions, peer-reviewed journals, and professional organizations (Chou et al., 2015; MedlinePlus, n.d.; UCSF Health, n.d.). To what extent these are the predominant sources in LLM-generated health responses—and to what extent they compete for source citation among commercial health and alternative sources—is the investigation of this study. Different from some other LLM models, such as ChatGPT and Perplexity, that use Microsoft Bing as their primary search engine (Klaine, 2025; Mehdi, 2023), Claude uses Brave Search as its search backend (Wiggers, 2025; Hatmaker, 2025), which may also influence the sources it retrieves.

Given these gaps, this research seeks to understand and establish a baseline for Claude's health-related citation behavior, the characteristics of the sources most likely to be cited by Claude AI. We do so by applying the Authority Signals Framework (Jacques et al., 2026), originally developed to analyze ChatGPT health citations, to identify the authority signal patterns present in Claude's cited sources. Understanding which sources AI systems prioritize requires asking four critical questions:

1. Who wrote it? (Author Credentials)
2. Who published it? (Institutional Affiliation)
3. How was it vetted? (Quality Assurance)
4. How does AI find it? (Digital Authority)

## 2. Methods

### 2.1 Data Source and Sample

This descriptive cross-sectional study used data from HealthSearchQA, which contains 3,172 consumer health questions curated by Google Research from publicly available search engine suggestions. These questions reflect common queries associated with medical search terms. After excluding 64 questions that received responses from Claude without cited sources and 33 questions with non-health-related citations, a final dataset of 3,075 questions was used for this study.

### 2.2 Data Collection and Sampling Procedures

#### 2.2.1 Phase 1: Citation Extraction

In February 2026, the 3,172 questions were stored in a Supabase PostgreSQL database and the Claude Sonnet 4.5 API was called for each question to generate source citations. That extraction yielded 12,006 URLs. Those 12,006 were subsequently scraped by Apify to retrieve source names and organizational types for each URL, written to a Google Sheet following an automated pipeline developed and stored on GitHub and hosted and deployed on Vercel. This initial data extraction was initiated using Node.js code with full details available via Jacques, 2026, Zenodo. Of those 12,006 URLs, source names and organizational types were successfully scraped from 10,111.

Of the 3,172 questions, 64 received responses from Claude without cited sources and were excluded. An examination of source names and organizational types led to the identification of questions and citations that were not health-related, including questions related to technology and questions that were inaccurately interpreted by Claude due to ambiguity (e.g., "What do flashes mean?" was interpreted as a weather question). Thus, 33 questions and their associated 73 citations were removed from the analysis, leaving a final sample of 3,075 questions yielding 10,038 citations.

A JavaScript-based validation tool deployed as a Google Sheets extension applied the same operational definitions used in the companion ChatGPT study (Jacques et al., 2026) to verify and correct automated classifications. Sources were classified into seven organization types: (1) Medical Institution, (2) Government Resource, (3) Commercial Health Information, (4) Professional Association & Advocacy, (5) Encyclopedia/Reference, (6) Peer-Reviewed Journal, and (7) News/Media. Complete operational definitions are available via Zenodo [https://doi.org/10.5281/zenodo.19616668].

#### 2.2.2 Phase 2: Sampling

A second phase employed a disproportionate stratified sampling strategy to further examine the 10 authority signals of those sources by organizational type. A Google Scripts code was developed using a Fisher-Yates shuffling algorithm to select a 5% sample from all organizational types that had more than 100 citations. The full census was used for those with fewer than 100. Additionally, the full census of Commercial Health Information was included given that it was the non-established source category. Among the disproportionate stratified sample, 296 had broken URLs in which the page was either moved or not operational, leaving a sample of n = 542: Medical Institution n = 130 (from 2,600 total in Phase 1 for that type), Government Resource n = 99 (from 1,980), Professional Association/Resource n = 100 (from 2,000), Encyclopedia n = 69, Peer-Reviewed Journal n = 34, News/Media n = 7, and Commercial Health Information n = 103.

### 2.2.3 Phase 3: Authority Signal Coding

The third phase analyzed the 10 authority signals of the 542 disproportionate stratified sample citations. An automated pipeline was developed using Node.js code containing: (1) the operational definitions of the 10 authority signals, (2) instructions to pull the HTML source code from each URL retrieved from Apify, (3) directions to package the HTML source code with the 10 authority signal definitions, and (4) instructions to send the definitions and HTML source code to Claude Sonnet 4.5 to analyze and write results onto Google Sheets. That code was stored on GitHub and hosted and deployed on Vercel. The Vercel URL endpoint was used to trigger the analyses, and automation was initiated using a Chrome extension that automatically refreshed the endpoint URL after each row was completed. Results were written to Google Sheets and updates or errors of the analyses were stored in Supabase.

Page Authority, Domain Authority, and Spam Score were collected via the Moz API. Three human coders (ED, AC, JD) reviewed all 542 sources across all 10 authority signal variables to verify the accuracy of the automated coding against the operational definitions. Where discrepancies were identified, coding decisions were reviewed and corrected.

### 2.2.4 Variables and Measures

The Authority Signals Framework developed in the companion ChatGPT study (Jacques et al., 2026) was applied to examine 10 authority signals across four domains for the disproportionate stratified sample:

1. Who wrote it? (Author Credentials)
2. Who published it? (Institutional Affiliation)
3. How was it vetted? (Quality Assurance)
4. How does AI find it? (Digital Authority)

**Who wrote it? (Author Credentials Domain)** includes one signal: (2) author attribution level, coded as 0 = none, 1 = name only, 2 = name with credentials.

**Who published it? (Institutional Affiliation Domain)** includes one signal: (1) organization type, classified into seven categories: Medical Institution, Government Resource, Commercial Health Information, Professional Association & Advocacy, Encyclopedia/Reference, Peer-Reviewed Journal, and News/Media.

**Figure 1: Data collection and analysis pipeline**

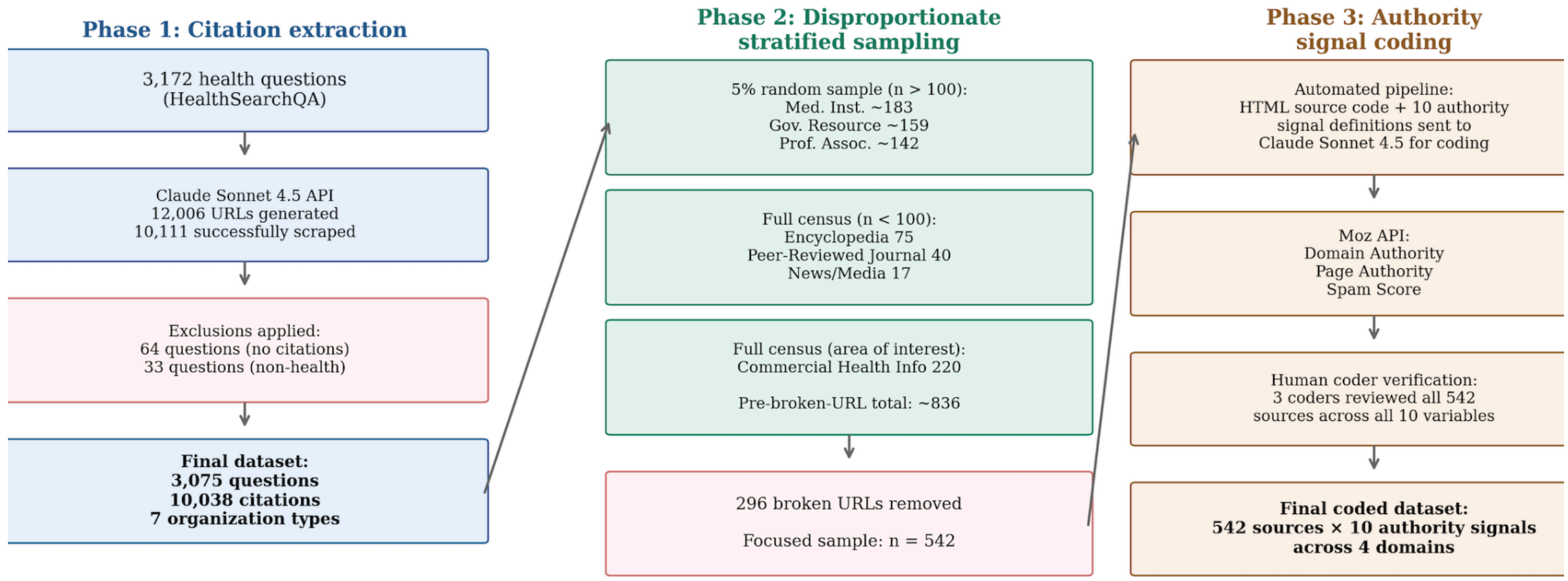

**How was it vetted? (Quality Assurance Domain** includes three signals: (3) references cited (yes/no); (4) medical review statement visible (yes/no); and (5) content recency, coded as 0 = before 2020 or not listed, 1 = 2020–2023, 2 = 2024–2026.

**How does AI find it? (Digital Authority Domain)** includes five signals: (6) page authority, (7) domain authority, and (8) spam score, each on 0–100 scales (collected via Moz API); (9) content length, coded as 0 = brief <500 words, 1 = moderate 500–1,500 words, 2 = comprehensive >1,500 words; and (10) schema markup, coded as 0 = none, 1 = microdata, 2 = JSON-LD.

Complete operational definitions for all variables are consistent with the companion study and available via Zenodo [https://doi.org/10.5281/zenodo.19616668].

2.2.5 Analysis

Descriptive statistics were calculated for the full citation dataset (N = 10,038), including frequencies and percentages for organization type distribution. For the disproportionate stratified sample (n = 542), descriptive statistics including frequencies, percentages, medians, and interquartile ranges were calculated for all authority signal variables. Authority signal characteristics were examined across organization types to identify patterns in credibility strategies. Software used included Google Sheets for data preparation and Python for statistical analysis.

## 3. Results

### 3.1 What Types of Sources Does Claude Cite?

Table 1 presents the organization type distribution of all 10,038 Claude health citations identified in Phase 1. Established institutional sources—Medical Institutions, Government Resources, Professional Associations, Encyclopedias, Peer-Reviewed Journals, and News/Media—accounted for 97.8% of all citations (n = 9,818). Commercial Health Information comprised the remaining 2.2% (n = 220). Medical Institutions were the most frequently cited organization type (n = 3,666, 36.5%), followed by Government Resources (n = 3,172, 31.6%) and Professional Associations & Advocacy (n = 2,848, 28.4%). The top 10 most frequently cited organizations accounted for 57.8% of all citations (n = 5,691), with Mayo Clinic alone representing 24.7% (n = 2,476) (Figure 2).

**Figure 2: Top 10 most frequently cited organizations in Claude health citations (N = 10,038**

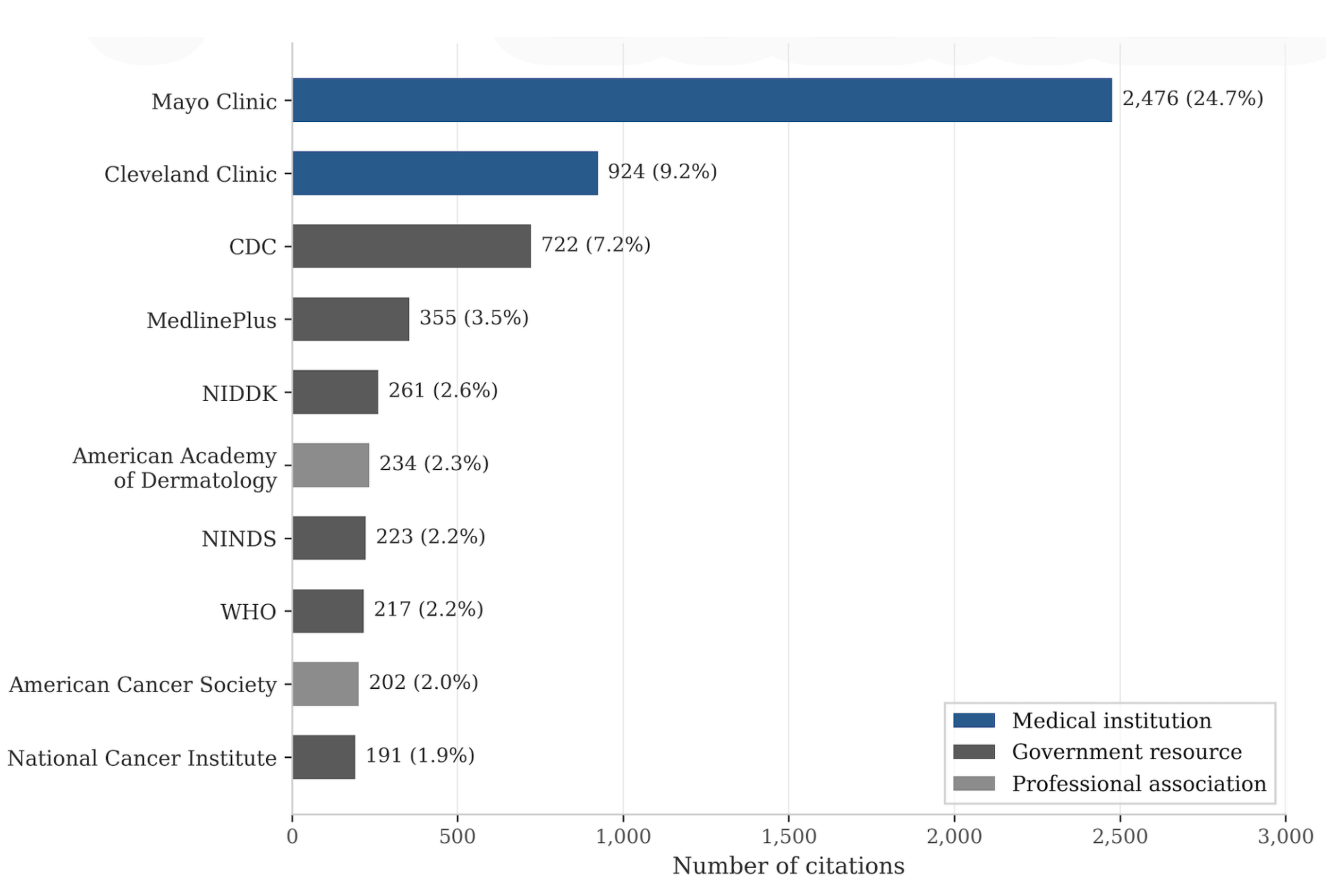

Note. Top 10 organizations = 57.8% of all 10,038 citations (n = 5,805).

**Table 1. Organization Type Distribution of Claude Health Citations (N = 10,038)**

| Organization Type | Total Citations | % | Analysis Sample | Sample Method |
|---|---|---|---|---|
| Medical Institutions | 3,665 | 36.5 | ~183 | 5% sample |
| Government Resources | 3,173 | 31.6 | ~159 | 5% sample |
| Professional Assoc. & Advocacy | 2,848 | 28.4 | ~142 | 5% sample |
| Commercial Health Info | 220 | 2.2 | 220 | Census |
| Encyclopedia/Reference | 75 | 0.7 | 75 | Census |
| Peer-Reviewed Journal | 40 | 0.4 | 40 | Census |
| News/Media | 17 | 0.2 | 17 | Census |
| Total | 10,038* | 100.0 | ~836 | |

**Note: Analysis Sample reflects the pre-broken-URL stratified sample (~836). After removing 296 broken URLs, the final focused analysis sample was n = 542. Rounding of individual percentages may result in minor discrepancies.*

### 3.2 What Credibility Characteristics Do Those Sources Have?

The disproportionate stratified analysis sample (n = 542) was constructed through stratified sampling with deliberate oversampling of Commercial Health Information (full census) to enable detailed comparison. The following results describe authority signal characteristics within this analytical sample and should not be interpreted as representative of Claude's overall citation distribution, which is reported in Table 1.

***Author Credentials.*** The majority of Claude-cited sources in the disproportionate stratified sample provided no author attribution (n = 395, 72.9%). Among sources with attribution, 15.9% (n = 86) provided a name only and 11.3% (n = 61) provided a name with credentials. Author attribution varied substantially by organization type: Medical Institutions had no author attribution (100%), while Commercial Health Information sources were most likely to provide named authors with credentials (Table 2).

***Institutional Affiliation.*** In the full citation dataset (N = 10,038), Claude overwhelmingly cited established institutional sources, which accounted for 97.8% of all citations (n = 9,818). Medical Institutions were the most frequently cited (n = 3,665, 36.5%), followed by Government Resources (n = 3,173, 31.6%), Professional Associations & Advocacy (n = 2,848, 28.4%), Commercial Health Information (n = 220, 2.2%), Encyclopedia/Reference (n = 75, 0.7%), Peer-Reviewed Journals (n = 40, 0.4%), and News/Media (n = 17, 0.2%). The Professional/Practice Website category, present in the companion ChatGPT study, was not represented in Claude citations. The top 10 organizations accounted for 57.8% of all citations (n = 5,805), with Mayo Clinic alone representing 24.7% (n = 2,476) (Figure 2).

***Quality Assurance.*** Medical review statements were present in 37.1% of disproportionate stratified sample sources (n = 201), references were cited in 28.0% (n = 152), and 56.6% of sources (n = 307) had content dated 2024–2026. Commercial Health Information sources had the highest rate of medical review statements (86.4%), exceeding all other organization types including Medical Institutions (44.6%) and Professional Associations (33.0%). Government Resources had the lowest medical review rate (14.1%). Peer-Reviewed Journals had the highest rate of references cited (76.5%) (Table 2).

***Digital Authority.*** Claude-cited sources in the disproportionate stratified sample had a median Domain Authority of 92 (IQR 82–93), median Page Authority of 61 (IQR 54–67), and median Spam Score of 3 (IQR 1–3). Schema markup was present in 67.5% of sources (n = 366), with JSON-LD as the predominant type (55.7%). The majority of sources were comprehensive in length (59.0%, >1,500 words). Medical Institutions had the highest schema markup adoption (92.3%) and comprehensive content rates (92.3%), while Government Resources had among the lowest schema markup rates (43.4%) despite having the highest median Domain Authority (95) (Table 2).

**Table 2. Authority Signals by Organization Type (n = 542)**

*Percentages shown for binary/categorical variables; medians shown for continuous variables.*

| Authority Signal | Med Inst | Gov | Prof Assn | Comm Hlth | Encyc | Peer-Rev | News | Total |
|---|---|---|---|---|---|---|---|---|
| n | 130 | 99 | 100 | 103 | 69 | 34 | 7 | 542 |
| Medical Review (%) | 44.6 | 14.1 | 33.0 | 86.4 | 7.2 | 5.9 | 0.0 | 37.1 |
| References Cited (%) | 16.2 | 39.4 | 23.0 | 35.0 | 8.7 | 76.5 | 14.3 | 28.0 |
| Schema Markup (%) | 92.3 | 43.4 | 45.0 | 82.5 | 72.5 | 50.0 | 85.7 | 67.5 |
| Comprehensive (%) | 92.3 | 29.3 | 57.0 | 71.8 | 20.3 | 70.6 | 28.6 | 59.0 |
| Recent 2024–26 (%) | 74.6 | 54.5 | 41.0 | 77.7 | 39.1 | 14.7 | 42.9 | 56.6 |
| No Author (%) | 100.0 | 89.9 | 81.0 | 23.3 | 91.3 | 14.7 | 42.9 | 72.9 |
| DA Median | 92 | 95 | 73 | 84 | 92 | 95 | 92 | 92 |
| PA Median | 66 | 67 | 54 | 60 | 57 | 60 | 58 | 61 |
| Spam Median | 3 | 1 | 1 | 3 | 3 | 1 | 1 | 3 |

## 4. Discussion

This study applied the Authority Signals Framework (Jacques et al., 2026) to establish a baseline for Claude AI's health-related citation behavior, drawing on 10,038 citations generated in response to consumer health questions. What the data shows is that Claude leans, and leans heavily, on established institutional sources. Those sources account for 97.8% of everything it cites. But that preference comes with a catch: the sources are concentrated among a small number of organizations. The top ten organizations account for 57.8% of all citations (n = 5,805), and Mayo Clinic by itself holds 24.7% (n = 2,476). Claude is drawing from traditionally credible categories, namely Medical Institutions (36.5%), Government Resources (31.6%), and Professional Associations (28.4%). Still, when a handful of organizations carry that much of the load, it is fair to ask how much diversity of health perspective actually survives the filter. Commercial Health Information sources made up just 2.2% of all citations (n = 220), which is small. What is inside that small slice, though, is interesting. The commercial sources Claude did cite were unusually dense with visible quality signals. 86.4% carried medical review statements, 82.5% used schema markup, and 71.8% ran past 1,500 words of content. Traditional institutional sources, by contrast, showed up in Claude's citations with or without any of those markers. Government Resources had the highest median Domain Authority in the set (95), yet the lowest medical review rate (14.1%) and a modest rate of schema adoption (43.4%). That same asymmetry suggests two different logics at work: institutional sources seem to get cited on name and authority alone, while commercial sources that make the cut tend to arrive wearing every credibility badge they own. Whether those signals are what resulted in commercial health sources being cited, or whether they are simply an association present among those selected, is something this study does not settle. That question would deserve its own study.

The digital fingerprint of the sources Claude cites (high Domain Authority with a median of 92, low Spam Scores with a median of 3, and schema markup on 67.5% of them) echoes search retrieval patterns more broadly. Claude runs on Brave Search (Wiggers, 2025), which is a different backend than ChatGPT's Microsoft Bing (Mehdi, 2023), and that difference may be part of what shapes the specific pattern we observed here, including the complete absence of Professional/Practice Website citations that did appear in the companion ChatGPT study (Jacques et al., 2026). Both models prefer institutional sources. The specific institutions they land on, and how tightly they concentrate around them, do not look the same. Citation behavior, it turns out, is not just a function of the model. The retrieval layer underneath matters too.

These findings sit alongside a growing peer-reviewed literature on where LLM-generated health information falls short. Wu et al. (2025) found that somewhere between 50% and 90% of LLM health responses are not fully supported by the sources they cite. Bedi et al. (2025), in their systematic review of 519 studies, pointed out that most LLM evaluations in healthcare are measuring accuracy and almost none are looking at the credibility of the sources themselves. This study takes up that exact gap. It does not ask whether Claude's citations are correct. It asks what authority signals describe the sources Claude chooses in the first place.

LLMs are becoming part of how people look for health information, not in theory but in practice. Recent data puts the share of users consulting an LLM chatbot for health information at over 21% (Yun & Bickmore, 2025). At the same time Anthropic is positioning Claude for HIPAA-ready healthcare applications (Anthropic, 2026). That combination is the reason a baseline matters now rather than later. This study provides that baseline. It also shows that the Authority Signals Framework travels across platforms, which means the same framework can be used for ongoing monitoring and for cross-model comparison as these systems keep moving deeper into clinical workflows. If we are going to let these tools into the room where care is happening, we need to know, in detail, what they are citing and why.

## 5. Limitations

Several limitations should be considered when interpreting these results. First, this study does not evaluate the factual accuracy of Claude’s citations; it examines the characteristics of the sources cited rather than whether those sources substantively supported Claude’s response. Second, the study design is descriptive and cross-sectional. Citation behavior may change over time as Claude’s backend, ranking algorithms, or healthcare integrations evolve. Third, the research used a disproportionate stratified sampling strategy (n = 542) that deliberately oversampled Commercial Health Information to enable detailed comparison, and was further reduced by 296 broken URLs. Because some organization types may have lost more URLs than others, both of these factors may affect representativeness within organization types. Fourth, analyses were limited to consumer-facing health questions from the HealthSearchQA dataset and may not generalize to clinical prompts or provider-directed queries.

Finally, the API was called using default parameters, including a temperature setting of 1.0, which allows for more randomness in responses. Given that over 10,000 citations were pulled, the likelihood that this would have affected the overall distribution patterns is minimal.

## 6. Conclusion

Claude’s health citation behavior reflects established institutional preference, with 97.8% of citations originating from non-commercial, traditionally credible categories, but this preference is concentrated around a small number of highly recognizable organizations. The authority signals present in cited sources vary substantially by organization type: traditional institutions are cited regardless of visible quality assurance markers, while the commercial sources that do appear tend to exhibit medical review statements, schema markup, and comprehensive content. These findings establish a baseline for Claude’s citation behavior as LLMs increasingly enter healthcare workflows, and demonstrate the utility of the Authority Signals Framework as a tool for ongoing, cross-platform evaluation of AI-mediated health information.

### Data Availability

All research study materials and coding pipeline are publicly available at: [https://doi.org/10.5281/zenodo.18287499](https://doi.org/10.5281/zenodo.18287499)

### AI Statement

AI was used to support data collection, literature review search, data verification, figure creation, and editorial review of manuscript drafts.

### Software and Tools

Moz. (2024). *Moz Domain Authority*. Moz Pro. Retrieved from [https://moz.com/learn/seo/domain-authority](https://moz.com/learn/seo/domain-authority)

Anthropic. (2024). *Claude API: Claude Sonnet 4* [Application Programming Interface]. Retrieved from https://www.anthropic.com/api

Apify Technologies. (2024). *Website Content Crawler* [Web scraping tool]. Retrieved from https://apify.com/apify/website-content-crawler

Google. (2024). *Google Sheets API* (V4) [Application Programming Interface]. https://developers.google.com/sheets/api

HealthSearchQA Dataset. openmedlab. "Awesome-Medical-Dataset/Resources/HealthSearchQA. Md at Main · Openmedlab/Awesome-Medical-Dataset." GitHub. Accessed April 17, 2026. [https://github.com/openmedlab/Awesome-Medical-Dataset/blob/main/resources/HealthSearchQA.md](https://github.com/openmedlab/Awesome-Medical-Dataset/blob/main/resources/HealthSearchQA.md).

Supabase Inc. (2024). *Supabase: The open source Firebase alternative* [Software]. Retrieved from https://supabase.com

## References

Anh-Hoang, D., Tran, V., & Nguyen, L.-M. (2025). Survey and analysis of hallucinations in large language models: Attribution to prompting strategies or model behavior. *Frontiers in Artificial Intelligence*, *8*. https://doi.org/10.3389/frai.2025.1622292

Anthropic. (2026, January 11). *Advancing Claude in healthcare and the life sciences*. https://www.anthropic.com/news/healthcare-life-sciences

Bedi, S., Liu, Y., Orr-Ewing, L., Dash, D., Koyejo, S., Callahan, A., Fries, J. A., Wornow, M., Swaminathan, A., Lehmann, L. S., & others. (2025). Testing and evaluation of health care applications of large language models: A systematic review. *Jama*, *333*(4), 319–328.

Chou, W., Gaysynsky, A., & Persoskie, A. (2015). Health literacy and communication in palliative care. *Textbook of Palliative Care Communication*, 90–101.

Fortune Editors. (2023). *Anthropic's CEO says why he quit his job at OpenAI to start a competitor that just received billions from Amazon and Google*. Yahoo Finance. https://finance.yahoo.com/news/anthropic-ceo-says-why-quit-194409797.html

Jacques, E. (2026). ejacques1/Claude-authority-signals: Authority Signals in Claude AI Health Citations: Code, Data, and Operational Definitions (v1.0). Zenodo. https://doi.org/10.5281/zenodo.19616668

Jacques, E., Datuowei, E., Jones, V., Basch, C., Vanderpool, C., Udeozo, N., & Chapa, G. (2026). *Authority Signals in AI Cited Health Sources: A Framework for Evaluating Source Credibility in ChatGPT Responses* (arXiv:2601.17109). arXiv. https://doi.org/10.48550/arXiv.2601.17109

Jin, H., Guo, J., Lin, Q., Wu, S., Hu, W., & Li, X. (2024). Comparative study of Claude 3.5-Sonnet and human physicians in generating discharge summaries for patients with renal insufficiency: Assessment of efficiency, accuracy, and quality. *Frontiers in Digital Health*, *6*, 1456911. https://doi.org/10.3389/fdgth.2024.1456911

Klaine, J. (2025, March 25). *Perplexity AI and the use of Bing | HyperLinker*. HyperLinker. https://www.hyperlinker.ai/en/seo/perplexity-ai-bing

Linardon, J., Jarman, H. K., McClure, Z., Anderson, C., Liu, C., & Messer, M. (2025). Influence of Topic Familiarity and Prompt Specificity on Citation Fabrication in Mental Health Research Using

Large Language Models: Experimental Study. *JMIR Mental Health*, *12*, e80371. https://doi.org/10.2196/80371

MedlinePlus. (n.d.). *Evaluating Health Information* [Text]. National Library of Medicine. Retrieved April 15, 2026, from https://medlineplus.gov/evaluatinghealthinformation.html

Mehdi, Y. (2023, February 7). Reinventing search with a new AI-powered Microsoft Bing and Edge, your copilot for the web. *The Official Microsoft Blog*. https://blogs.microsoft.com/blog/2023/02/07/reinventing-search-with-a-new-ai-powered-microsoft-bing-and-edge-your-copilot-for-the-web/

Szkutak, R. (2025, July 31). Enterprises prefer Anthropic's AI models over anyone else's, including OpenAI's. *TechCrunch*. https://techcrunch.com/2025/07/31/enterprises-prefer-anthropics-ai-models-over-anyone-elses-including-openais/

UCSF Health. (n.d.). *Evaluating Health Information*. Ucsfhealth.Org. Retrieved April 15, 2026, from https://www.ucsfhealth.org/health-articles/evaluating-health-information

Wiggers, K. (2025, March 21). Anthropic appears to be using Brave to power web search for its Claude chatbot. *TechCrunch*. https://techcrunch.com/2025/03/21/anthropic-appears-to-be-using-brave-to-power-web-searches-for-its-claude-chatbot/

Wu, K., Wu, E., Wei, K., Zhang, A., Casasola, A., Nguyen, T., Riantawan, S., Shi, P., Ho, D., & Zou, J. (2025). An automated framework for assessing how well LLMs cite relevant medical references. *Nature Communications*, *16*(1), 3615. https://doi.org/10.1038/s41467-025-58551-6

Yun, H. S., & Bickmore, T. (2025). Online health information–seeking in the era of large language models: Cross-sectional web-based survey study. *Journal of Medical Internet Research*, *27*, e68560.